\documentclass[preprint,10pt,a4paper,prd]{revtex4}
\usepackage{graphicx}
\usepackage{amsmath}
\newcommand{\dint}{\int}

\begin{document}
\title{The UV behavior of Gravity at Large N}
\author{F. Canfora}
\email[]{canfora@sa.infn.it}
%\bigskip
%\begin{frontmatter}
\affiliation
{Istituto Nazionale di Fisica Nucleare, Sezione di Napoli, GC di Salerno\\
Dipartimento di Fisica ''E.R.Caianiello'', Universit\`{a} di
Salerno\\ Via S.Allende, 84081 Baronissi (Salerno), Italy}

\begin{abstract}
A first step in the analysis of the renormalizability of gravity
at Large N is carried on. Suitable resummations of planar diagrams
give rise to a theory in which there is only a finite number of
primitive superficially divergent Feynman diagrams. The mechanism is
similar to the the one which makes renormalizable the 3D Gross-Neveu model
 at large N. The connections with gravitational
confinement and KLT relations are shortly analyzed. Some potential
 problems in fulfilling the Zinn-Justin equations are pointed out.
\end{abstract}
%\begin{keyword}
%Large N expansion, Einstein-Hilbert action, Holographic Principle.
%\PACS: 11.15.Pg, 04.90.+e, 04.50.+h, 11.10.Gh.
%\end{keyword}

%\end{frontmatter}
\maketitle
\section{Introduction}

\noindent The quantum theory of renormalizable interactions plays the main
role in our present understanding of fundamental physical laws. It gave rise
to the formulation of the Standard Model of elementary particles which is
still today in a very good shape. Of course, there are many problems which
have not been solved yet (such as confinement in QCD): some of these
problems are likely to be "technical" problems in the sense that a better
understanding of the actual standard theory should be enough to give the
correct solutions. As far as other problems (such as the hierarchy problem,
the cosmological constant, the quantum version of the gravitational
interaction and so on) are concerned, the standard theory is likely to be
inadequate. In particular, the lacking of a precise understanding of non
perturbative phenomena occurring in the strongly coupled phase of gravity
(which should clarify important and still poorly understood features of
early cosmology) is unpleasant: being gravity perturbatively non
renormalizable, it is not possible to make detailed predictions in such a
phase. The two main candidates to be the final theory of quantum gravity,
\textit{Superstring Theory} and \textit{Loop Quantum Gravity} (which are not
necessarily to be thought as mutually exclusive), still are too complicated
to be fully understood. Thus, it is worth to explore new ways in which
"physical effects beyond the standard model" could be already manifest in
the low energy (low with respect to the Planck scale) physics. The dominant
point of view of string theorists is to see the Einstein-Hilbert action as
an effective action in which the "heavy degrees of freedom" have been
integrated out. Unfortunately, in order to make this idea predictive and to
clarify how it is possible to improve in practice the UV behavior of gravity
it would be important to understand in more details the dynamics of such
degrees of freedom and, at the present stage, this is a rather difficult
task.

It is not enough to say that "gravity is an effective field
theory\textquotedblright : even if this is the case, to perform meaningful
computations in the strongly coupled regime with the technical tools at our
disposal (perturbative expansions of various kind, renormalization group and
so on), physical effects \textquotedblright beyond the standard
model\textquotedblright\ (related, for instance, to string theory) which
enable such a meaningful (and, hopefully, predictive) computations still
have to be clarified. To be more specific, string theory predicts various
type of geometrical corrections to the \textquotedblright
bare\textquotedblright\ Einstein-Hilbert action:
\begin{equation*}
S_{corr}=S_{EH}+\sum_{n}\left( \alpha ^{\prime }\right)
^{n}\int_{M}f_{n}(R_{\mu \nu \rho \sigma })d\mu
\end{equation*}%
where $S_{EH}$ is the Einstein-Hilbert action, $\alpha ^{\prime }$ is the
string length and the $f_{n}$ are higher order curvature invariants. Even if
a large but finite number of terms are added, the corrected action remains
non renormalizable and, consequently, it is not yet clear how to perform
meaningful quantum computations (unless one would be able to sum the whole
series: a hopeless task indeed). In a sense, the analysis of the $\alpha
^{\prime }$ corrections as a tool to shed light on non perturbative
phenomena in gravity is like the analysis of the standard perturbation
expansion in QCD to understand confinement. Clearly, in QCD, there is a
little hope to understand non perturbative phenomena with ordinary
perturbation theory. In the same way, in gravity one should expect physics
beyond the standard model to manifest itself in a different, perhaps more
subtle, way. Renormalizability is not a mere aesthetic requirement; it is,
in fact, the need to have a theory which is predictive in the strongly
coupled phase: one should expect (it is better to say \textquotedblright
hope\textquotedblright ) that physics beyond the standard model will improve
the Einstein-Hilbert gravity precisely in this direction.

A sound theoretical idea which could give important indications on the way
to follow is the \textit{Holographic Principle} introduced in \cite%
{Su95,tH93} (for two detailed reviews see \cite{AG00a,Bo02}) which is at the
basis of the string-theoretical AdS/CFT correspondence \cite{Ma97}. In a
recent paper \cite{Ca05}\ a Large \textbf{N} expansion for the gravitational
interaction has been formulated which shed new light on the relations
between higher spins, the holographic principle and non perturbative
phenomena such as (a sort of) gravitational confinement. In \textbf{SU(N)}\
\textit{Gauge Theory}, the large \textbf{N} expansion introduced by 't Hooft
in \cite{T74a,T74b} and refined by Veneziano \cite{Ve76} is indeed one of
the most powerful non perturbative techniques available to investigate the
strongly coupled phase (it provides the issues of \textit{confinement},
\textit{chiral symmetry breaking} and the relation with \textit{string theory%
} with a rather detailed understanding; a clear analysis of the role of
Baryons at large \textbf{N} has been given in \cite{Wi79}: see, for two
detailed pedagogical reviews, \cite{Ma98}).

One of the main properties of the large \textbf{N} expansion is that many
non trivial models (such as the $O(N)$ $\phi ^{4}$ model in five space-time
dimension and the Gross-Neveu model in three space-time dimensions which are
not renormalizable in the standard perturbative expansion) become in fact
renormalizable (see, for example, \cite{Pa75}): the Green functions are not
anymore analytic in the small coupling constant region (see, for example,
\cite{FP63,Lee62,RS60}). Thus, the large \textbf{N} expansion can be seen as
a strong coupling expansion which, besides to clarify non perturbative
phenomena such as \textit{confinement} and \textit{chiral symmetry breaking}%
, greatly improve the UV behavior of theories which, at a first glance,
would appear meaningless at high energies. Here, it will be argued that,
under very reasonable assumptions, this could also happen in gravity. The
large \textbf{N} expansion suggests suitable resummations of planar diagrams
which lead to an UV-softening of gravity: the relation between the Newton
constant and the mass of higher spin field(s) seems to be quite similar to
the relation between the Fermi coupling constant and the mass of the $W_{\pm
}$ bosons of the electro-weak interactions.

The paper is organized as follows: in section 2 the diagrammatic formulation
of General Relativity as a constrained BF theory is shortly described and
the well known result about the perturbative non renormalizability of
gravity is described in this formalism. In section 3, the large \textbf{N}
resummations which improve the UV behavior of gravity are introduced: for
the sake of clarity, in section 3 the complications connected to ghosts will
be neglected keeping manifest, however, the main physical features leading
to the UV softening. In section 4, the ghosts effects which could prevent
the UV softening are analyzed. In section 5, a possible physical
interpretation of the large \textbf{N} \textquotedblright UV
softening\textquotedblright\ is proposed and the connections with the KLT
relations equation are pointed out. Eventually, the conclusions are drawn.

\section{BF-gravity and Feynman rules}

In this section the BF formulation of gravity \ and the corresponding
Feynman rules will be shortly described.

The topological \textit{BF} theory\footnote{%
Topological, in this setting, means that the theory has no local degrees of
freedom and the expectation values which can be computed are related to
topological invariants of the manifold where the theory lives.} in four
dimensions is defined by the following action
\begin{align}
S\left[ A,B\right] & =\dint\limits_{M}B^{IJ}\wedge ^{\ast }\left(
F_{IJ}(A)\right) =\frac{1}{4}\dint\limits_{M}\varepsilon ^{\alpha \beta
\gamma \delta }B_{\alpha \beta }^{IJ}F_{\gamma \delta IJ}d^{4}x  \label{bfa}
\\
B^{IJ}& =\frac{1}{2}B_{\alpha \beta }^{IJ}dx^{\alpha }\wedge dx^{\beta
},\quad F_{IJ}=\frac{1}{2}F_{\alpha \beta IJ}dx^{\alpha }\wedge dx^{\beta }
\notag \\
F_{\alpha \beta IJ}& =\left( \partial _{\alpha }A_{\beta }-\partial _{\beta
}A_{\alpha }\right) _{IJ}+A_{\alpha I}^{L}A_{\beta LJ}-A_{\beta
I}^{L}A_{\alpha LJ},  \label{fieldstr}
\end{align}
where $M$ is the four-dimensional space-time, $\ast $ is the Hodge dual, the
greek letters denote space-times indices, $\varepsilon ^{\alpha \beta \gamma
\delta }$\ is the totally skew-symmetric Levi-Civita symbol in
four-dimensional space-times, $I$, $J$ and $K$ are the internal Lorentz
indices which are raised and lowered with the Minkowski metric $\eta _{IJ}$:
$I,J=1,..,\mathbf{N}$. Thus, the basic fields are a $so(\mathbf{N}-1,1)$%
-valued differential 2-form $B_{IJ}$ and a $so(\mathbf{N}-1,1)$ connection
1-form $A_{\alpha LJ}$, the internal gauge group being $SO(\mathbf{N}-1,1)$.
Also the Riemannian theory can be considered in which the internal gauge
group is $SO(\mathbf{N})$ and the internal indices are raised and lowered
with the Euclidean metric $\delta _{IJ}$; in any case, both $B_{IJ}$ and $%
A_{\alpha LJ}$ are in the adjoint representation of the (algebra of the)
internal gauge group. The equations of motion are
\begin{eqnarray}
F &=&0,\quad \nabla _{A}B=0,  \label{vabf} \\
\nabla _{A} &=&\nabla =d+\left[ A,.\right]  \notag
\end{eqnarray}
where $\nabla _{A}$ is the covariant derivative with respect to the
connection $A_{\alpha LJ}$. The above equations tell that $A_{\alpha LJ}$
is, locally, a pure gauge and $B^{IJ}$\ is covariantly constant. When $%
\mathbf{N}=4$ and $B^{IJ}$ has the form

\begin{equation}
B^{IJ}=\frac{1}{2}\varepsilon _{KL}^{IJ}e^{K}\wedge e^{L}.  \label{preco}
\end{equation}
The action (\ref{bfa}) is nothing but the Palatini form of Einstein-Hilbert
action. Eq. (\ref{preco}) can be enforced by adding to the action (\ref{bfa}%
) a suitable constraint: the basic action in the BF formalism is

\begin{equation}
GS_{GR}=S\left[ A,B\right] -\dint\limits_{M}\left( \phi _{IJKL}B^{IJ}\wedge
B^{KL}+\mu H\left( \phi \right) \right) ,  \label{gra}
\end{equation}
where $G$ is the gravitational coupling constant, $\mu $ is fixed a
differential 4-form and $H(\phi )$ is a scalar which may have one of the
following expressions:
\begin{equation}
H_{1}=\phi _{IJ}^{IJ},\quad H_{2}=\phi _{IJKL}\varepsilon ^{IJKL},\quad
H_{3}=a_{1}H_{1}+a_{2}H_{2},  \label{enfo1}
\end{equation}
the $a_{i}$ being real constants (see \cite{Ca91,Ca91b,Ca01,De99,Re99}). It
is worth to note here that the Lagrange multiplier $\phi $ has four internal
indices. The form (\ref{gra}) of the Einstein-Hilbert action is a natural
starting point to formulate the ''gravitational''\ large \textbf{N}
expansion \cite{Ca05} since the connection formulation allows to adopt the
double line notation (however, being the fundamental representation of $so(%
\mathbf{N}-1,1)$ real, the lines of internal indices carry no arrows).

The classical action in Eq. (\ref{gra}) is left invariant by $so(\mathbf{N}%
-1,1)-$gauge transformations and by diffeomorphisms. The analysis of the
BRST invariance of the gravitational action in the BF formalism can be found
in \cite{LN93}. As far as the scope of the present paper is concerned, a
technical complication is that the natural kinetic term (see Eq. (\ref{prop1}%
) below) is invariant under a further transformation
\begin{equation*}
\delta _{3}A_{\mu }=0,\quad \delta _{3}B_{\mu \nu }=\nabla _{\left[ \mu
\right. }\omega _{\left. \nu \right] }
\end{equation*}
(where $\omega _{\nu }$\ is a $so(\mathbf{N}-1,1)$ valued 1-form) which has
to be gauge fixed too in order to derive the propagators (a detailed
discussion of this issue can be found in \cite{FMST97}, \cite{Cat98}).

The most suitable way to proceed is to follow \cite{Cat98} in which the
authors (in the case of the BF formulation of Yang-Mills theory) introduced
an auxiliary non physical field $\eta _{\mu }$ (a Lie algebra valued
1-form), in the combination
\begin{equation*}
B^{\prime }=B-\nabla \eta ,
\end{equation*}%
whose role is to keep at the same time both the local degrees of freedom of
Yang-Mills theory and the symmetries of the $BF$ theory (physically, $\eta $%
\ represents the longitudinal components of $B$). It is worth to note here
that, because of the Bianchi identities, one has
\begin{equation*}
S\left[ A,B^{\prime }\right] =S\left[ A,B\right] .
\end{equation*}%
In the Yang-Mills case, this procedure is lawful as is\ because the
(classical) action of the $BF$ Yang-Mills theory is
\begin{equation}
S_{YM}^{cl}=S\left[ A,B\right] -e^{2}\dint\limits_{M}tr\left( B^{\prime
}\right) _{\mu \nu }\left( B^{\prime }\right) ^{\mu \nu }  \label{YMuno}
\end{equation}%
(where $e$ is the Yang-Mills coupling constant). In gravity the second term
on the right hand side of the above equation is replaced by the constraint
in Eq. (\ref{gra})
\begin{equation*}
\dint\limits_{M}\left( \phi _{IJKL}B^{IJ}\wedge B^{KL}+\mu H\left( \phi
\right) \right) .
\end{equation*}%
The consequence is that, in order to produce the Yang-Mills kinetic term for
$\eta $, it is convenient to add the second term on the right hand side of
Eq. (\ref{YMuno}) to the gravitational action. This is possible since it has
been shown in \cite{FMST97} \cite{Cat98} that the small $e$ limit manifests
no problem (the theory is \textquotedblright perturbative\textquotedblright\
in $e$). In other words, the second term on the right hand side of Eq. (\ref%
{YMuno}) can be regarded as a true vertex: therefore one can consider the
gravitational case as the small $e$ limit of the action\footnote{%
It is worth to stress here that the results of the present paper, and in
particular the large \textbf{N} improvement of the UV behavior of gravity,
do not depend on $e$. The second term on the right hand side of Eq. (\ref%
{YMuno}) has been added only to obtain a theory with the same propagators as
the BF Yang-Mills theory: this scheme allows to clearly single out the
\textquotedblright bad UV-behaved\textquotedblright\ vertex of the theory.
At this stage of the analysis it is not yet clear if it is possible to get
the same UV improvement without adding the above term to the gravitational
action.} (\ref{0gracla}).

Indeed, $\eta $\ has only a technical role since it\ just represents the
longitudinal components of $B$ and one of the $\eta $'s transformation laws
is given by a shift (see below Eq. (\ref{2sym})). However, at this stage of
the analysis, it seems to be unavoidable to add the Yang-Mills term (the
second term on the right hand side of Eq. (\ref{YMuno})) in order to obtain
the Feynman rules, vertices and propagators. On the other hand, due to the
comparison with Yang-Mills theory, this scheme is mandatory if one wants to
clearly identify the "guilty of the perturbative non renormalizability of
gravity" in the BF scheme: but for this enlargement of the gravitational
action, it would not be possible to "large \textbf{N}" improve the UV
behavior of gravity.

The classical symmetries of the \textquotedblright
enlarged\textquotedblright\ classical $BF$ gravitational action
\begin{equation}
GS_{cl}=S\left[ A,B\right] -\dint\limits_{M}\left( \phi _{ab}\left(
B^{\prime }\right) ^{a}\wedge \left( B^{\prime }\right) ^{b}+\mu H\left(
\phi \right) \right) -e^{2}\dint\limits_{M}tr\left( B^{\prime }\right)
^{a}\left( B^{\prime }\right) _{a}  \label{0gracla}
\end{equation}%
(where $a$, $b$, $c$ and so on\ are indices in the adjoint representation of
$so(\mathbf{N}-1,1)$) are
\begin{eqnarray}
\delta _{1}A_{\mu } &=&\left( \nabla \theta _{(1)}\right) _{\mu };\quad
\delta _{1}B_{\mu \nu }=\left[ B_{\mu \nu },\theta _{(1)}\right] ;\quad
\delta _{1}\eta _{\mu }=\left[ \eta _{\mu },\theta _{(1)}\right] ;
\label{1sym} \\
\delta _{2}A_{\mu } &=&0;\quad \delta _{2}B_{\mu \nu }=\nabla _{\left[ \mu
\right. }\omega _{\left. \nu \right] };\quad \delta _{2}\eta _{\mu }=\omega
_{\mu }  \label{2sym} \\
\delta _{3}A_{\mu } &=&0;\quad \delta _{3}B_{\mu \nu }=\left[ F_{\mu \nu
},\theta _{(3)}\right] ;\quad \delta _{3}\eta _{\mu }=\nabla _{\mu }\theta
_{(3)}  \label{3sym}
\end{eqnarray}%
where $\theta _{(i)}$\ are $so(\mathbf{N}-1,1)$ valued gauge scalars. $%
\delta _{1}$ is a simple gauge transformation so that the action is
invariant. As far as $\delta _{2}$ and $\delta _{3}$ are concerned, the
transformations of $B$ cancel out the transformations of $\eta $ in the
second and in the third terms on the right hand side of Eq. (\ref{0gracla}).
The BF term is left invariant by $\delta _{1}$ because of the Bianchi
identities and by $\delta _{2}$ because it reduces to a trivial gauge
transformation on the usual $F^{2}$ term of (the standard formulation of)
Yang-Mills theory. It is worth to note that the symmetries $\delta _{2}$ and
$\delta _{3}$ are reducible, as it is clear if one considers in Eqs. (\ref%
{2sym}) and (\ref{3sym})
\begin{equation*}
\left( \nabla \theta _{(3)}\right) _{\mu }=\omega _{\mu }.
\end{equation*}%
To obtain the Feynman rules, the gauge fixing and ghost terms related to the
above symmetries have to be included. A convenient gauge fixing is
\begin{equation}
\partial _{\mu }A^{\mu }=0,\quad \partial _{\mu }\eta ^{\mu }=0,\quad
\partial _{\mu }B^{\mu \nu }=0  \label{gf1}
\end{equation}%
so that the corresponding gauge-fixing termof the action is
\begin{eqnarray}
S_{gf} &=&\dint\limits_{M}\left\{ \overline{c}\left( -\partial _{\mu }\nabla
^{\mu }\right) c+h_{A}\left( \partial _{\mu }A^{\mu }\right) +\right.  \notag
\\
&&+\overline{\psi }^{\nu }\partial ^{\mu }\left\{ -\left[ B_{\mu \nu },c%
\right] +\nabla _{\left[ \mu \right. }\psi _{\left. \nu \right] }+\left[
F_{\mu \nu },\rho \right] \right\} +  \notag \\
&&+h_{B}\left( \partial _{\mu }B^{\mu \nu }\right) +\overline{\rho }\partial
^{\mu }\left\{ -\left[ \eta _{\mu },c\right] +\psi _{\mu }+\nabla _{\mu
}\rho \right\} +  \notag \\
&&+h_{\eta }\left( \partial _{\mu }\eta ^{\mu }\right) +u\left( \partial
^{\mu }h_{B}\right) +h_{\overline{\psi }}\left( \partial _{\nu }\overline{%
\psi }^{\nu }\right) +  \notag \\
&&\left. +\overline{\xi }\partial ^{\mu }\left\{ \left[ \psi _{\mu },c\right]
+\nabla _{\mu }\xi \right\} +h_{\psi }\left( \partial _{\nu }\psi ^{\nu
}\right) \right\} ,  \label{fixthegauge}
\end{eqnarray}%
where $\left( c,\overline{c},h_{A}\right) $, $\left( \psi ,\overline{\psi }%
,h_{B}\right) $ and $\left( \rho ,\overline{\rho },h_{\eta }\right) $ are
respectively the ghost, the antighost and the Lagrange multiplier for $%
\delta _{1}$ and $\delta _{2}$; $\left( \xi ,\overline{\xi },h_{\psi
}\right) $ the ghost, the antighost and the Lagrange multiplier for the zero
modes of the topological symmetry $\delta _{3}$ and $\left( u,h_{\overline{%
\psi }}\right) $ a pair of fields which takes into account a further
degeneracy associated with $\overline{\psi }$. It is worth to stress here
that all the fields appearing in the gauge fixing term (\ref{fixthegauge})
are in the adjoint representation of the gauge group: this will be important
when the role of ghosts loops effects in not preventing the UV softening
will be discussed. The following tables summarize the ghost numbers and
dimensions of the fields
\begin{eqnarray*}
&&%
\begin{array}{cccccccccc}
Fields & A & B & \eta & c & \overline{c} & \psi & \overline{\psi } & h_{A} &
h_{B} \\
dimension & 1 & 2 & 1 & 0 & 2 & 1 & 1 & 2 & 1 \\
ghost\quad number & 0 & 0 & 0 & 1 & -1 & 1 & -1 & 0 & 0%
\end{array}%
, \\
&&%
\begin{array}{ccccccccc}
Fields & \xi & \overline{\xi } & h_{\psi } & \rho & \overline{\rho } &
h_{\eta } & u & h_{\overline{\psi }} \\
dimension & 0 & 2 & 2 & 0 & 2 & 2 & 2 & 2 \\
ghost\quad number & 2 & -2 & -1 & 1 & -1 & 0 & 0 & 1%
\end{array}%
.
\end{eqnarray*}

The natural choice is to consider, as the Gaussian part of the fields $A$
and $B$, the off-diagonal kinetic term

\begin{equation}
S_{0}=\frac{1}{\kappa }\dint\limits_{M}\left( \varepsilon ^{\alpha \beta
\gamma \delta }B_{\alpha \beta }^{a}\partial _{\gamma }A_{\delta a}\right) ,
\label{prop1}
\end{equation}%
plus the quadratic terms for ghosts in the gauge fixing term (\ref%
{fixthegauge}). The $A\rightarrow B$ propagator (which propagates $A_{\mu }$
into $B_{\nu \gamma }$) has the following structure (a simple method to find
them can be found in \cite{Ho88} \cite{Mar97}):
\begin{equation}
\Delta _{(A,B)\mu \nu \gamma }^{(a,b)}=-\delta ^{ab}\frac{1}{2}\varepsilon
_{\mu \nu \gamma \alpha }\frac{p^{\alpha }}{p^{2}}.  \label{propag1}
\end{equation}%
The internal index structures of the propagators tells that, as one should
expect, the internal index is conserved along the gravitational internal
color lines. The $A\rightarrow A$, $B\rightarrow B$ and the $\eta
\rightarrow \eta $ propagators are
\begin{align}
\Delta _{(A,A)\mu \nu }^{(a,b)}& =\delta ^{ab}\frac{1}{p^{2}}(\delta ^{\mu
\nu }-\frac{p^{\mu }p^{\nu }}{p^{2}}),\quad \Delta _{(B,B)\mu \nu \gamma
\rho }^{(a,b)}=-\delta ^{ab}\varepsilon _{\mu \nu \alpha \lambda
}\varepsilon _{\gamma \rho \beta \lambda }\frac{p^{\alpha }p^{\beta }}{p^{2}}
\label{propaga4} \\
\Delta _{(\eta ,\eta )}^{(a,b)}& =(\delta ^{\mu \nu }-\frac{p^{\mu }p^{\nu }%
}{p^{2}})\delta ^{ab}.  \label{etapro}
\end{align}%
The ghosts propagators can be deduces from the gauge-fixing term in Eq. (\ref%
{fixthegauge}):
\begin{eqnarray*}
\Delta _{(\psi ,\overline{\psi })\mu \nu }^{(a,b)} &=&-i\delta ^{ab}\frac{1}{%
p^{2}}(\delta ^{\mu \nu }-\frac{p^{\mu }p^{\nu }}{p^{2}}),\quad \Delta _{(c,%
\overline{c})}^{(a,b)}=-i\delta ^{ab}\frac{1}{p^{2}} \\
\Delta _{(\xi ,\overline{\xi })}^{(a,b)} &=&-i\delta ^{ab}\frac{1}{p^{2}}%
,\quad \Delta _{(\rho ,\overline{\rho })}^{(a,b)}=-i\delta ^{ab}\frac{1}{%
p^{2}}.
\end{eqnarray*}%
Graphically, all the propagators in the double line notation are represented
by two parallel internal "gravitational" color lines (along which the
internal $so(\mathbf{N}-1,1)$ index is conserved) with no arrows \cite{T74a}
\cite{Ve76}.

\begin{figure}[tbp]
\begin{center}
\includegraphics*[scale=.30,angle=0]{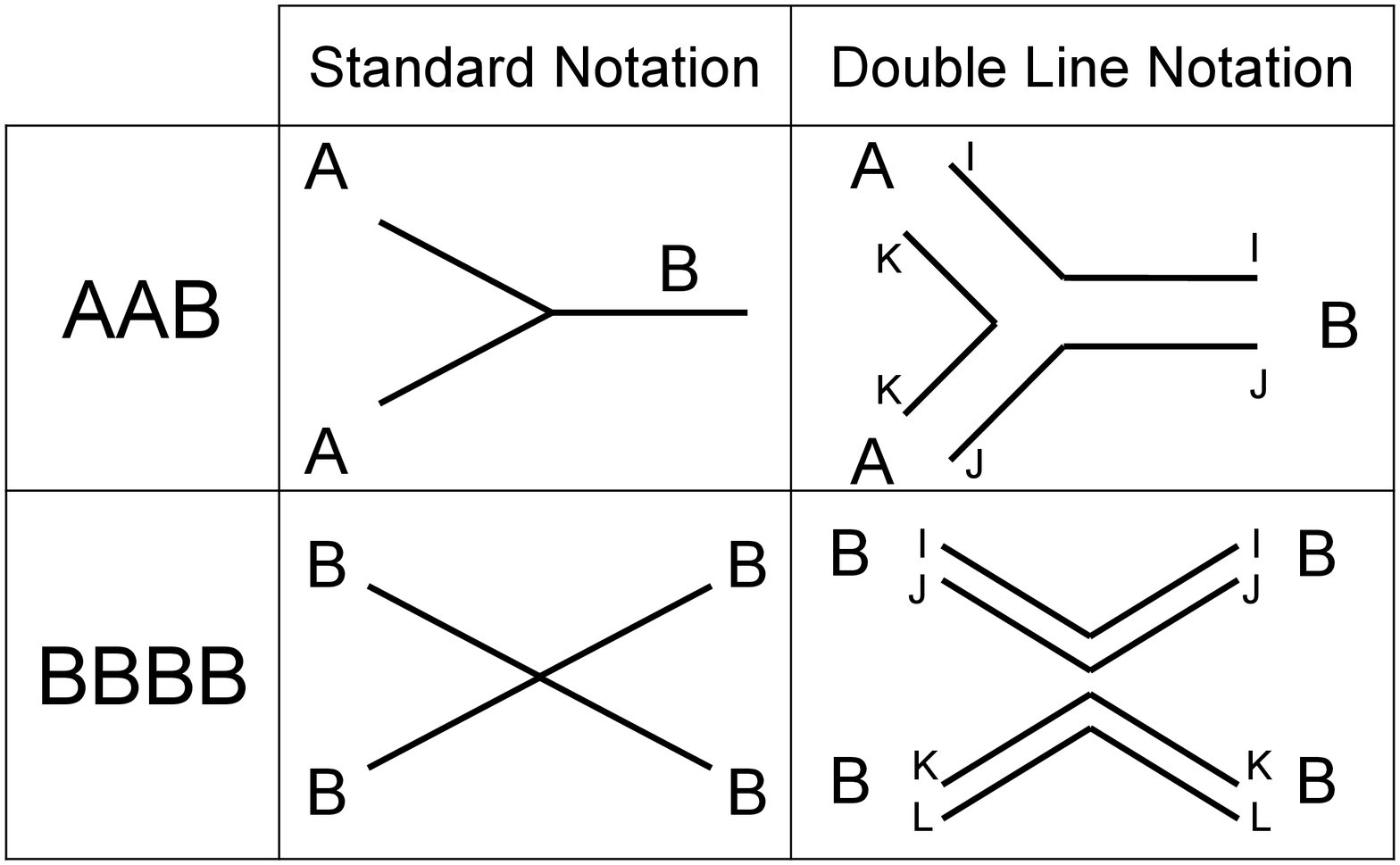}
\end{center}
\caption{Here, the double-line structures of the two matter vertices have
been displayed: the $AAB$ vertices is in common with the BF-Yang-Mills
theory, the 4-uple $B$-vertex is peculiar of gravity.}
\label{1}
\end{figure}

The coupling with matter fields (as discussed in \cite{Ca05}) shows that the
number of internal \textquotedblright gravitational color\textquotedblright\
lines associated with each matter field is connected with its spin: the
higher the spin, the more internal lines are needed (that is, the higher the
representation of the internal gauge group $SO(\mathbf{N}-1,1)$ is) to
describe the matter field in the double line notation (so that the Lagrange
multiplier field $\phi $ should be considered as a non propagating higher
spin field). For the sake of simplicity, in this paper the purely
gravitational case will be considered, however the inclusion of matter
fields should not destroy the main conclusions since, usually, matter fields
do not worsen the UV behavior.

\begin{figure}[tbp]
\begin{center}
\includegraphics*[scale=.30,angle=0]{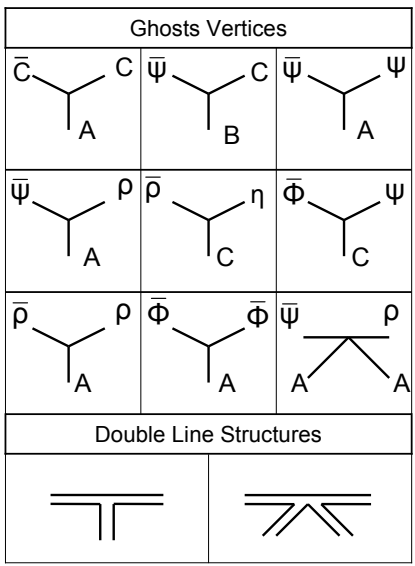}
\end{center}
\caption{In this picture the ghosts vertices have been drawn: they have only
two types of connected internal index structures.}
\label{2}
\end{figure}

The theory has the following matter vertices:

\begin{align}
V_{1}(A_{\mu }^{a},A_{\nu }^{b},B_{\alpha \beta }^{c})& =Gf^{abc}\varepsilon
_{\mu \nu \alpha \beta },\quad  \label{vertex} \\
V_{2}(B_{\mu \nu }^{a},B_{\alpha \beta }^{b},\phi ^{cd})& =G\delta
_{ac}\delta _{bd}\varepsilon ^{\mu \nu \alpha \beta },  \label{vertex2}
\end{align}%
where $f^{abc}$ are the structure constants (the Newton constant $G$, as
usual, has been absorbed in the fields in such a way that it appears only in
the vertices which are depicted in Fig. \ref{1}). The first vertex is also
present in the BF formulation of Yang-Mills theory while the second one only
pertains to General Relativity and is likely to be the main responsible for
the quantum realization of the holographic principle \cite{Ca05}. The ghosts
vertices in the gauge fixing term are:
\begin{eqnarray}
V_{3}(A_{\mu }^{a},c^{b},\overline{c}^{d}) &=&-Gf^{abd}p^{\mu },\quad V_{4}(%
\overline{\psi }^{a\sigma },B_{\gamma \nu }^{b},c^{d})=-Gf^{abd}p^{\nu
}\delta _{\sigma }^{\gamma },  \label{vert34} \\
\frac{V_{5}(A_{\mu }^{a},\psi _{\nu }^{b},\overline{\psi }_{\sigma }^{d})}{G}
&=&-f^{abd}p^{\left[ \mu \right. }\delta ^{\left. \nu \right] \sigma },\quad
\frac{V_{6}(A_{\mu }^{a},\rho ^{b},\overline{\psi }_{\sigma }^{d})}{G}%
=f^{abd}p^{\left[ \mu \right. }\delta ^{\left. \nu \right] \sigma }p_{\nu },
\label{vert56} \\
V_{7}(A_{\mu }^{a},\rho ^{b},\overline{\rho }^{d}) &=&-Gf^{abd}p^{\mu
},\quad V_{8}(c^{a},\eta _{\nu }^{b},\overline{\rho }^{d})=Gf^{abd}%
\varepsilon _{\mu \nu \alpha \beta },  \label{vert78} \\
V_{9}(A_{\mu }^{a},\xi ^{b},\overline{\xi }^{d}) &=&Gf^{abd}p^{\mu },\quad
V_{10}(c^{a},\psi _{\mu }^{b},\overline{\xi }^{d})=Gf^{abd}p^{\mu }
\label{vert910} \\
V_{11}(A_{\omega }^{m},A_{\tau }^{n},\rho ^{b},\overline{\psi }_{\sigma
}^{d}) &=&G^{2}f^{abd}p^{\left[ \mu \right. }\delta ^{\left. \nu \right]
\sigma }\delta _{\left[ \nu \right. }^{\tau }\delta _{\left. \mu \right]
}^{\omega }f^{mna}.  \label{vertu11}
\end{eqnarray}%
It is worth to note that the internal index structures of the ghosts
vertices is standard, that is, they have a connected structure (see, Fig.~\ref{2}%
): this fact will plays an important role in the following.

The aim of this paper is to show that, in the large \textbf{N} expansion,
there is only a finite number of superficially divergent diagrams. However,
large \textbf{N} resummations (even if they improve the UV behavior of the
theory) could give rise to some problems in fulfilling the Zinn-Justin
equation which is needed to ensure that the infinities can be absorbed in
counterterms not violating the symmetries of the original action. This point
will be discussed in slightly more details later on.

\subsection{The "non renormalizable" vertex}

The perturbative non renormalizability of (super)gravity \cite{GS86} was an
important result: even if at a first glance this could be rather obvious by
power counting, there are many examples (such as gravity in three dimensions
\cite{Wi88}) of theories which are trivially renormalizable (being
BRS-exact) and, in fact, would not appear in this way by naive power
counting arguments. Moreover, the results in \cite{TV74} about the one-loop
finiteness of the Einstein-Hilbert action lead to the expectations that the
powerful symmetries of gravity could give rise to some \textquotedblright
miracle\textquotedblright\ at least in supergravity: indeed, in the standard
perturbative formulation, such a miracle does not occur. It is important to
understand this result in the present formalism; the question is: which is
the \textquotedblright wrong\textquotedblright\ vertex responsible for the
perturbative non renormalizability of the theory? The answer is as follows:
if one would drop the vertex in Eq. (\ref{vertex2}), one would obtain the
topological BF Yang-Mills theory which, obviously, is renormalizable. Thus,
the vertex responsible for the perturbative non renormalizability of the
theory is the one in Eq. (\ref{vertex2}). In gravity, the problem is mainly
connected to the 4-uple $B$ vertex. Indeed, $B$ has a bad asymptotic
behavior (see Eq. (\ref{propaga4})): such a propagator is also present in
the BF Yang-Mills theory \cite{Mar97}. However, in the Yang-Mills case the $%
B $ field is always attached to the better-behaved $A$ field through the
good $AAB$ vertex. Consequently, loops with only $B$ fields cannot arise.
Such loops in which only the $B$ fields appear are at the origin of the
perturbative non renormalizability of gravity. The 4-uple vertex for $B$
(see the left column of Fig.\ref{5} in which there is a typical
non-renormalizable diagram which cannot arise in the BF formulation of
Yang-Mills theory) gives rise to bad UV-behaved loops in which only $B$%
-fields appear: at any order in perturbation theory, new diagrams diverging
in the UV appear in the expansion.

\section{Large N resummations and effective propagators}

Here, it will be shown that the large \textbf{N} expansion suggests a useful
resummation of a certain class of planar diagrams which greatly improves the
UV behavior of gravity. For the sake of clarity, it will be firstly
presented a simplified version of the argument\footnote{%
Simplified here means that the possible complications due to ghosts effects
in the loops (which, in principle, could prevent the UV softening) are not
taken into account.} (in which, however, the main physical ideas are
manifest) leading to the UV improvement. In the next section, the effects of
ghosts in the loops will be discussed.

\begin{figure}[tbp]
\begin{center}
\includegraphics*[scale=.30,angle=-90]{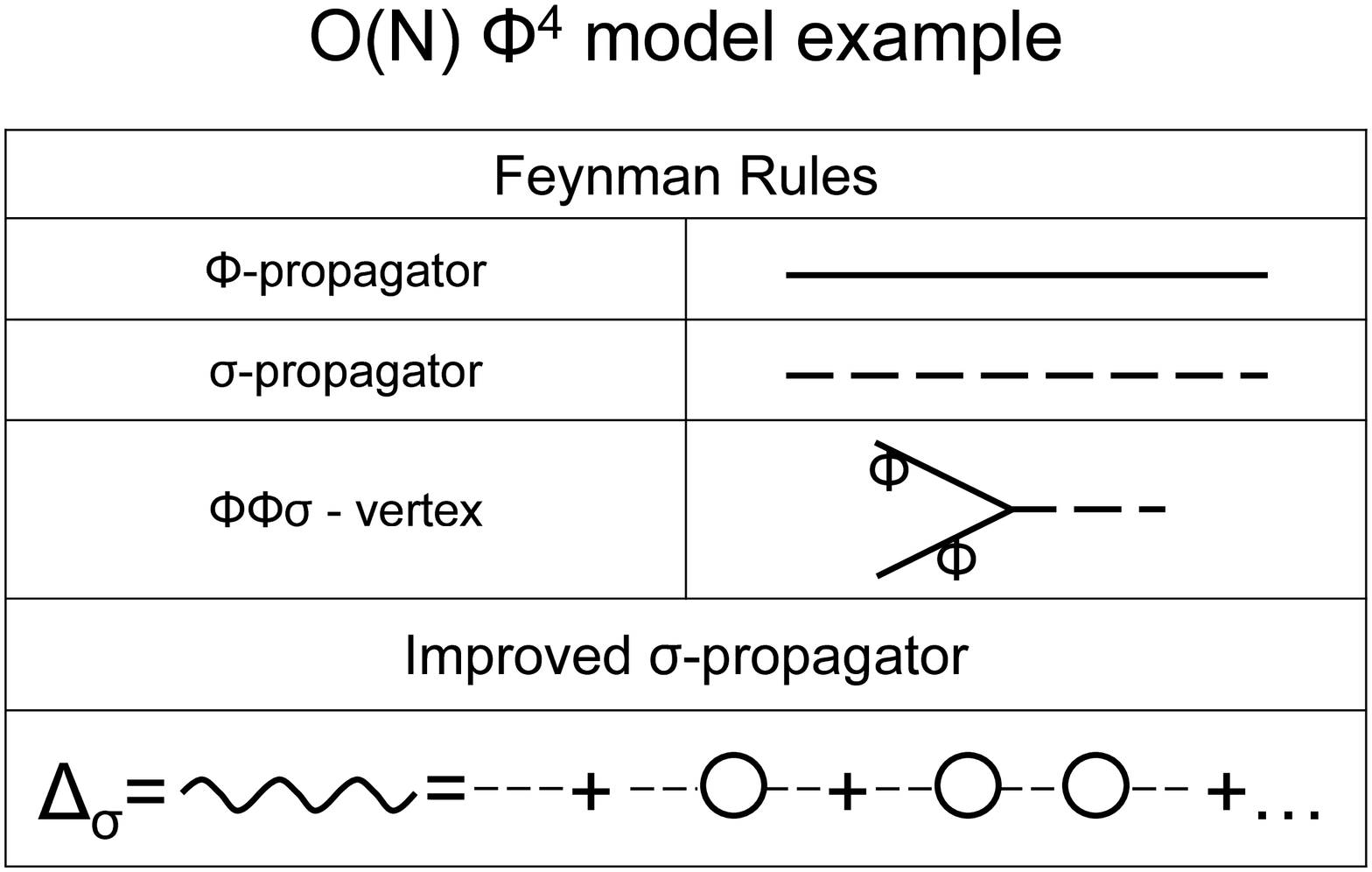}
\end{center}
\caption{{}In this picture the large \textbf{N} improvement of the scalar $%
\protect\phi ^{4}-O(N)$ model is depicted: the constant propagator of the
Lagrange multiplier is UV-softened since the leading large \textbf{N}
contribution is given by the geometric sum of bubble diagrams.}
\label{3}
\end{figure}

It is worth to briefly recall what happens in the large \textbf{N} expansion
of the $5D$ $O(\mathbf{N})$ $\phi ^{4}$ model \cite{Pa75}. The Lagrangian is
\begin{equation}
L_{\phi }=\frac{1}{2}\left( \partial _{\mu }\phi _{j}\partial ^{\mu }\phi
^{j}+m^{2}\phi _{j}\phi ^{j}\right) +\frac{\lambda }{8}\left( \phi _{j}\phi
^{j}\right) ^{2},\quad j=1,..,\mathbf{N}  \label{ac1}
\end{equation}%
$\phi _{j}$ being an $N-$component scalar field, $\lambda $ the coupling
constant and $m$ the mass. It is convenient to introduce a Lagrange
multiplier field $\sigma $ as follows
\begin{equation}
L_{\phi ,\sigma }=\frac{1}{2}\left( \partial _{\mu }\phi _{j}\partial ^{\mu
}\phi ^{j}+\left( m^{2}+i\sigma \sqrt{\lambda }\right) \phi _{j}\phi
^{j}\right) +\frac{1}{2}\sigma ^{2},  \label{ac2}
\end{equation}%
of course, due to the equation of motion of $\sigma $, this Lagrangian is
equivalent to the previous one. In the momentum space, the bare propagator $%
D_{\sigma }(k)$ of $\sigma $ is constant:
\begin{equation*}
D_{\sigma }(k)=1.
\end{equation*}%
The Lagrangian in Eq. (\ref{ac2}) in five space-time dimensions gives rise
to a non renormalizable theory. On the other hand, by the taking into
account that at large \textbf{N} the dominating diagrams are the well known
\textit{bubble diagrams}, it is possible to obtain an improved propagator $%
\overline{D}_{\sigma }(k)$ at the leading order in $1/\mathbf{N}$ (see Fig.%
\ref{3}):

\begin{equation*}
\overline{D}_{\sigma }(k)=\frac{1}{1+g\Pi (k)},\quad \Pi (k)\underset{%
k\rightarrow \infty }{\rightarrow }k,\quad g=\lambda \mathbf{N}.
\end{equation*}
Such an effective propagator improves the UV behavior of the theory in such
a way that there are only three superficially divergent diagrams and the
theory becomes renormalizable in the $1/\mathbf{N}$\ expansion. This
\textquotedblright miracle\textquotedblright\ happens in the following way:
many diverging diagrams in the standard perturbative expansion (which are of
different order in the standard coupling constant $\lambda $) are, in fact,
of the same order in $1/\mathbf{N}$: this fact tells that such diagrams
should be summed together (such a resummation is easy being a geometric
series). The difference with respect the standard perturbative
renormalizability is that the propagator is not anymore analytic in the
effective coupling constant $g$ in the region of small $g$ (see, for
instance, \cite{FP63,Lee62,RS60}).

As far as gravity is concerned, one has to cure mainly the
\textquotedblright non renormalizable\textquotedblright\ vertex in Eq. (\ref%
{vertex2}). In the scalar $O(\mathbf{N})$ case, the large \textbf{N}
expansion tells that a suitable class of diverging diagrams (the "bubble
diagrams") should be summed together: of course, from the technical point of
view, the gravitational case is more difficult. However, similarities in the
two cases are indeed present. It is easy to see that the dominant
contributions to the 4-uple $B$ vertex come from bubble-like diagrams which,
form the internal index point of view, are \textquotedblright tree
diagrams\textquotedblright\ (see Fig.\ref{4}): that is, they are planar
diagrams with no closed color loops so that they contribute to the "large
\textbf{N}-bare" vertex which is the relevant quantity as far as power
counting arguments are concerned\footnote{%
From a large \textbf{N} perspective, it is mandatory to sum together these
diagrams because, from the internal index point of view, they can be
considered on perfectly equal footing.}. Thus, one is simply computing the
tree-like term of the topological expansion at genus zero. These \textit{%
large-\textbf{N}-tree-diagrams} can be formally summed as geometric series
as it will be shown in a moment.

This is very much in the spirit of the large \textbf{N} expansion \cite%
{T74a,Ve76}: at any fixed genus, the contributions are weighted by the
effective coupling constant
\begin{equation}
g_{eff}=G\mathbf{N}  \label{geff}
\end{equation}%
(kept fixed at large \textbf{N}) to the power of the number of closed color
loops $L$ and, consequently, the tree diagrams should be summed together as
it happens in standard Feynman diagrams calculations. To be more precise, at
any given genus in the topological expansion, the large \textbf{N} diagrams
are weighted by the effective coupling constant in Eq. (\ref{geff}). To any
closed color loop in the topological expansion it corresponds a factor $%
g_{eff}$ in the same way as the powers of the Planck constant weight the
usual loops in the standard Feynman diagrams. Any interesting physical
observable $\left\langle O\right\rangle $ can be expanded at large \textbf{N}
as follows:
\begin{equation}
\left\langle O\right\rangle =\sum_{g,b}\mathbf{N}^{2-2g-b}\sum_{L}\left(
g_{eff}\right) ^{L}O_{g,b,L}  \label{lnstandard}
\end{equation}%
where $g$ is the genus, $b$ is the number closed matter loops, $L$ is the
number of closed color loops and $O_{g,b,L}$ is the sum of fat diagrams
contributing at genus $g$, with $b$ matter loops and $L$ closed color loops.
In the same way, in the standard Feynman expansion for physical quantities
such as amplitudes one has
\begin{equation*}
\left\langle O\right\rangle _{(F)}=\sum_{L_{F}}\left( \hbar \right)
^{L_{F}}O_{L_{F}}^{(F)}
\end{equation*}%
where $\hbar $\ is the Planck constant, $L_{F}$ is the number of loops and $%
O_{L_{F}}^{(F)}$ is the sum of the Feynman diagrams contributing at $L_{F}$
loops: the bare propagators and vertices have, by definition, $L_{F}=0$ (it
is apparent the similarity between the topological expansion with $g$ and $b$
fixed and the standard Feynman expansion). Thus, in very much the same way,
in the large \textbf{N} expansion the bare propagators and vertices must not
contain any closed color loops\footnote{%
In the Yang-Mills case the bare propagators and vertices in the large
\textbf{N} expansion coincide with the the bare propagators in the standard
Feynman expansion. Needless to say, the two expansions are, in any case,
very different.}. The key point is that to count the number of primitive
superficially divergent diagrams, one only needs tree-like propagtors and
vertices (see, for instance, \cite{We96}).

\subsubsection{UV-softening of the 4-uple $B$-vertex}

\begin{figure}[tbp]
\begin{center}
\includegraphics*[scale=.30,angle=0]{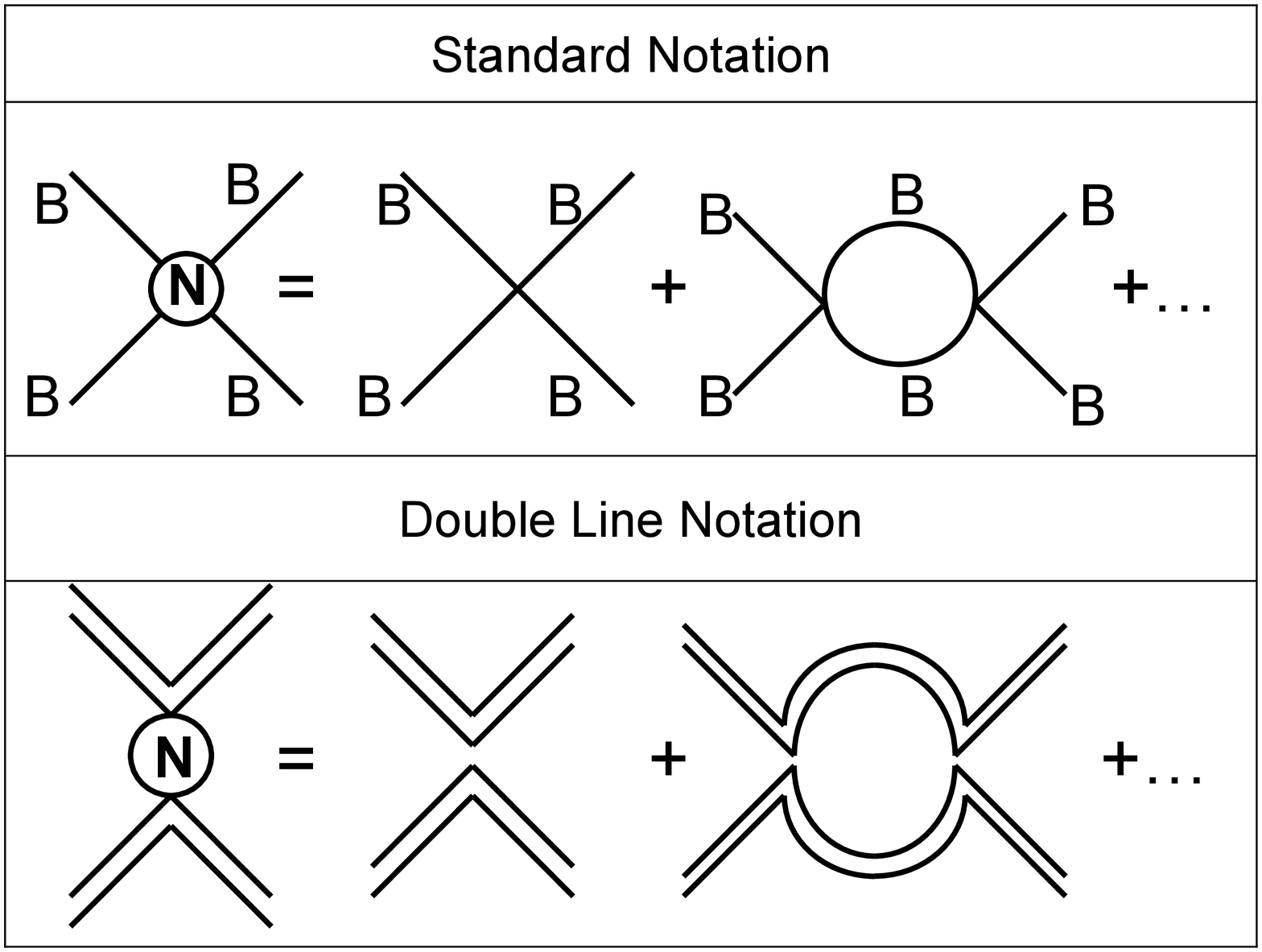}
\end{center}
\caption{{}In this picture it has been drawn the \textit{tree level large}
\textbf{N} 4-uple $B$ vertex. Because of the presence of the Lagrange
multiplier with four internal color indices, it is clear that the \textit{%
tree level large} \textbf{N} 4-uple $B$ vertex is the (geometric) sum of an
infinite number of terms which have no closed color loop but have an always
increasing number of loops of the standard Feynman expansion. Indeed, this
phenomenon is peculiar of gravity.}
\label{4}
\end{figure}

The "guilty of perturbative non-renormalizability" 4-uple $B$ vertex is UV
softened by large \textbf{N} effects: the factor which dresses $V_{2}$ is
related to the geometric "bubble-like" series in Fig. \ref{4}. It is worth
to stress here that the above "bubble-like" series is made of "tree-like
large \textbf{N}" diagrans: that is, it is the sum of diagrams with no
closed internal color lines ($L=0$). It is a peculiar feature of gravity the
possibility to construct a "tree-like large \textbf{N}" quantity (which,
therefore, enters at the leading order in the large \textbf{N} expansion)
which, in fact, contains an infinite number of standard Feynman loops. In
the large \textbf{N} expansion of Yang-Mills theory the "tree-like large
\textbf{N}" vertices and propagators coincide with the standard tree-like
Feynman vertices and propagators since the vertices have a connected
structures.

In the standard Feynman expansion the basic building blocks are tree-like
vertices and propagators (that is, vertices and propagators without Feynman
loops): starting from them, well known arguments (and, in particular, the
BPHZ theorem) tell whether or not the perturbation expansion has a finite
number of primitive superficially divergent diagrams. In very much the same
way, the analogous question in the large \textbf{N} expansion have to be
answered by looking at the "tree-like large \textbf{N}" vertices and
propagators\footnote{%
In the Yang-Mills case both the expansions (Feynman and large \textbf{N})
are trivially power-counting renormalizable.} (as it is clear from well
studied $O(\mathbf{N})$-vectorial examples: see, for instance, \cite{Pa75}).
Thus, to understand whether or not gravity in the large \textbf{N} expansion
has a finite number of primitive superficially divergent diagrams one has to
use "tree-like large \textbf{N}" vertices and propagators: thus the 4-uple $%
B $ vertex (which is a "tree-like quantity in the Feynman expansion) has to
be replaced by the "bubble-like" series in Fig. \ref{4} which correctly
accounts for all the tree-like large \textbf{N} contributions.

\begin{figure}[tbp]
\begin{center}
\includegraphics*[scale=.30,angle=0]{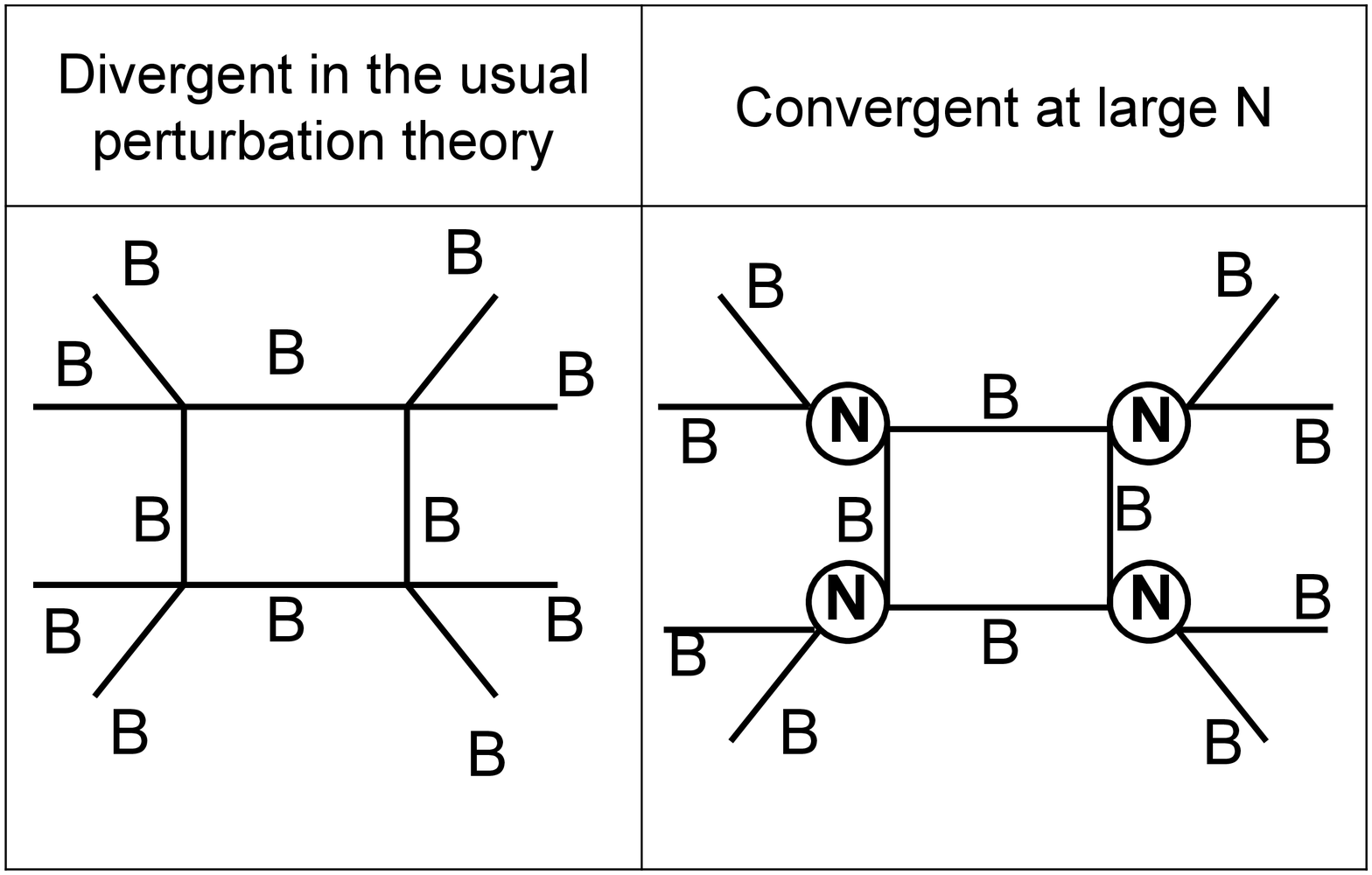}
\end{center}
\caption{{}In this picture it has been shown a typical diagram which is
divergent in the usual Feynman expansion and is, in fact, convergent at
large \textbf{N}. It is manifest that three "large \textbf{N} improved"
4-uple $B$-vertices are enough to make UV-finite a loop integral. This
should be compared with standard loops integrals with only $B$ propagators
which trigger the perturbative non-renormalizability of gravity: such loops
are not anymore a problem in the large \textbf{N} expansion.}
\label{5}
\end{figure}

Any potentially dangerous 4-uple $B$ vertex is dressed by a factor vanishing
in the UV as $1/q^{2}$. In Fig. \ref{5} is depicted a diagrams which
diverges in the standard perturbative expansion and is finite in the large
\textbf{N} improved expansion. This UV improvement appears to be quite
consistent with the Weinberg \textit{asymptotic safety scenario} \cite{We79}%
. The large \textbf{N} dressed vertex is
\begin{equation}
\mathbf{V}_{4B}(q)=V_{2}\frac{1}{1-\Pi (q^{2})},  \label{impropaga}
\end{equation}%
where $\Pi (q^{2})$ is the $B$ self-energy giving rise to the geometric
series:
\begin{eqnarray}
\Pi (q^{2}) &=&G^{2}\int_{\Lambda }d^{4}p\Delta _{(B,B)}(p)\Delta
_{(B,B)}(p-q)=  \label{ab111} \\
&=&G^{2}\left( \varepsilon ^{4}\right) _{\alpha \beta }^{\xi \eta
}\int_{\Lambda }d^{4}p\frac{p^{\alpha }p^{\beta }(p-q)_{\xi }(p-q)_{\eta }}{%
p^{2}(p-q)^{2}}\underset{\Lambda \rightarrow \infty }{\rightarrow }  \notag
\\
&&\underset{\Lambda \rightarrow \infty }{\rightarrow }\left(
I_{1}G^{2}\Lambda ^{2}\right) q^{2}+\left( G^{2}I_{2}\Lambda ^{4}\right) +...
\label{ab1111} \\
\left( \varepsilon ^{4}\right) _{\alpha \beta }^{\xi \eta } &=&\varepsilon
_{\mu \nu \alpha \lambda }\varepsilon _{\gamma \rho \beta \lambda
}\varepsilon ^{\mu \nu \xi \chi }\varepsilon ^{\gamma \rho \eta \chi },
\notag
\end{eqnarray}%
where $\Delta _{(B,B)}$ is the $B$ propagator, in the momentum integrals it
has been introduced a cutoff $\Lambda $ (to be removed in a suitable way),
terms which are subleading for $\Lambda \rightarrow \infty $ and \textbf{N}$%
\rightarrow \infty $ have been neglected and $I_{1}$ and $I_{2}$ are two non
vanishing real constants (whose precise values are not important as far as
the present discussion is concerned so that, from now on, they will be set
equal to one). It is worth to note that, in the geometric series giving rise
to $\Pi (q^{2})$, no factor \textbf{N} appears since there are no closed
color loops ($\Pi (q^{2})$ is a sum of the large \textbf{N} tree diagrams).
It is useful to rewrite Eq. (\ref{impropaga})
\begin{eqnarray}
\mathbf{V}_{4B}(q) &=&V_{2}\frac{M^{2}}{M_{0}^{2}-q^{2}+...},\quad
\label{effcost} \\
M_{0}^{2} &=&M^{2}+\delta M_{0}^{2}=M^{2}-b\Lambda ^{2},\quad M^{2}=\frac{1}{%
(G\Lambda )^{2}},  \notag
\end{eqnarray}%
where in the denominator of Eq. (\ref{effcost}) subleading terms when $%
\Lambda \rightarrow \infty $ have been neglected\footnote{%
One should also sum over cyclic permutations of the internal indices as well
as over the Mandelstam variables in orther to preserve crossing symmetry.}.
The most convenient way to remove the cutoff is
\begin{equation}
\frac{\Lambda }{\mathbf{N}}\underset{\Lambda ,\mathbf{N}\rightarrow \infty }{%
\rightarrow }finite\text{ }value\Rightarrow M^{2}=\frac{1}{(G\Lambda )^{2}}%
\rightarrow fixed,  \label{massfi}
\end{equation}%
where $M$ could be interpreted as a sort of renormalized Planck mass. The
divergent term $\delta M_{0}^{2}\sim -b\Lambda ^{2}$ has the typical form
which can be removed by a tadpole contribution.

\begin{figure}[tbp]
\begin{center}
\includegraphics*[scale=.30,angle=0]{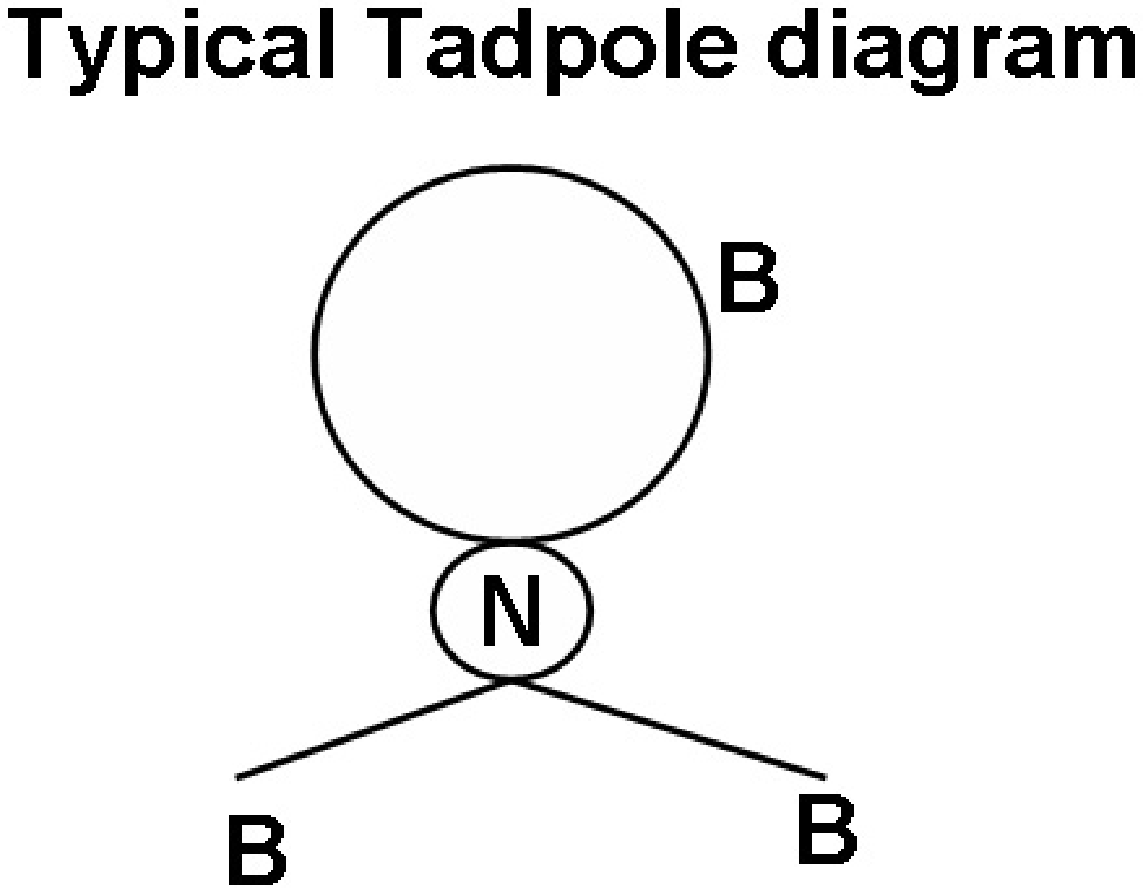}
\end{center}
\caption{
In this picture the typical tadpole diagram has been drawn. In
simpler models (such as the scalar $\protect\phi ^{4}-O(N)$ model) one can
take care of the tadpoles by using the so called \textit{gap equation}. In
gravity, it seems difficult to write and solve the gap equation so that one
has to remove by hand the tyical tadpole quadratic divergences.}
\label{8}
\end{figure}

The same phenomenon also occurs in simpler models in which the large \textbf{%
N} technique works (see, for instance, \cite{MZ03}). In such models the
quadratically divergent contribution to the mass is already included in the
so called \textit{gap equation}. The gap equation accounts for the tadpole
embodying it in the effective coupling constant(s). However, to do this, it
is necessary to write down and solve the saddle point equations at large
\textbf{N} for the full Lagrangian. In the gravitational case this seems to
be rather difficult so that one is enforced to remove the divergent
contribution to the mass "by hand" with counterterms. The typical tadpole
diagram\footnote{%
Of course, the tadpole diagrams also form a geometric series.} contributing
to the quadratically divergent contribution to the mass is in Fig. \ref{8}:
one can easily see that it is of order $\Lambda ^{2}$ due to the UV behavior
of the $B$ propagator.

Once the quadratically divergent term in the denominator is removed\footnote{%
If one would not remove such a term (which appears in the denominator of the
improved vertex), the improved $4$-uple $B$ vertex would actually vanish
when removing the cut-off. Therefore, one would obtain that the
gravitational action would be equal to the large \textbf{N} Yang-Mills
action in the limit of small coupling constant. Even if this would produce a
renormalizable theory in the large \textbf{N} expansion, it seems more
natural to remove the quadratically divergent term in the denominator as it
is usually done in similar situations \cite{MZ03}.}, one gets
\begin{equation}
\mathbf{V}_{4B}(q)=V_{2}\frac{M^{2}}{M^{2}-q^{2}+...}  \label{ultimpropro1}
\end{equation}%
where in the denominator terms which are subleading in the large \textbf{N}
limit have been neglected.
\begin{figure}[tbp]
\begin{center}
\includegraphics*[scale=.30,angle=0]{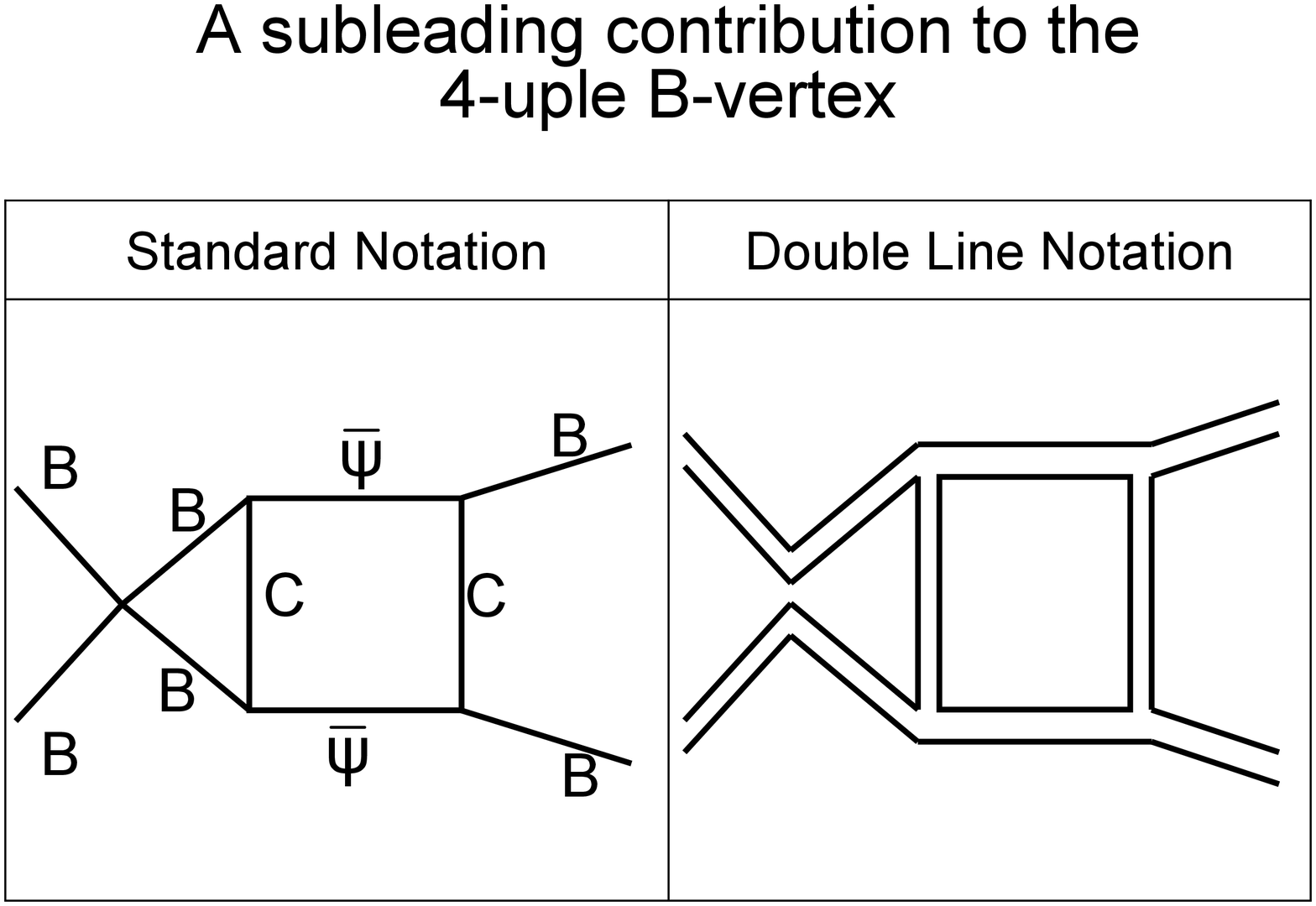}
\end{center}
\caption{
In this picture it has been drawn a typical ghosts contribution
to the 4-uple $B$ vertex. It is clear that, in the large \textbf{N}
expansion, the ghosts contributions are always subleading since, due to
their connected structures, they contain at least one closed color loop
(consequently, they do not contribute to the \textit{tree level large}
\textbf{N} 4-uple $B$ vertex).
}
\label{6b}
\end{figure}

It is now possible to compute the superficial degree of divergence of the
large \textbf{N} diagrams in which one simply has to use the 4-uple $B$
vertex in Eq. (\ref{ultimpropro1}) instead of $V_{2}$. Since any 4-uple $B$
vertex now decreases by 2 the superficially degree of divergence $\Omega $
of a loop integral, 3 vertex in Eq. (\ref{ultimpropro1}) are enough to make
a loop integral convergent: thus $\Omega \leq 4-E$ (where $E$ is the number
of external legs in the diagrams) as in the $BF$-Yang-Mills case \cite{Mar97}
(in Fig. \ref{5} there is meaningful example). The large \textbf{N}
prescription gives rise to a gravitational theory with only a finite number
of primitive superficially divergent diagrams (in good agreement with the
Weinberg \textit{asymptotic safety scenario} \cite{We79}): indeed, this is a
rather non-trivial step since it opens the unexpected possibility to control
the UV behavior gravity exploring, for instance, the inflationary phase of
cosmology.

However, in theories with local symmetries, one has also to show that the
infinities of the theory have the suitable symmetries which allow to cancel
such infinities by adding counter-terms not violating the original
symmetries of the action (see, for instance, \cite{We96}). It is not clear
at this stage of the analysis if such \textquotedblright large \textbf{N}
reorganization\textquotedblright\ prevents the fulfillment of the
Zinn-Justin equation: for instance, the large \textbf{N} expansion could not
commute with the BRS and Slavnov operators. This is a rather involved
technical question: I hope to return on this issue in a future pubblication.

It would be very interesting to compare more closely the above
results with the ones obtained in a series of papers \cite{Smo82}
\cite{Re97} \cite{Re97b} \cite{Re97c} \cite{Li03} in which the
authors began the analysis of the Einstein-Hilbert in the standard
metric variables using the method of the \textit{non perturbative
renormalization group}. The authors found sound evidences
supporting the existence of a non trivial UV fixed point (which,
obviously, would confirm the Weinberg \textit{asymptotic safety
scenario} \cite{We79}). The present results, although derived in a
completely different way, seem to be consistent with such a
scheme.

The above UV-improvement is conceptually very similar to what happens in the
change from the Fermi effective model of weak interaction to the
Glashow-Weinberg-Salam model of electro-weak interactions: in this case the
non renormalizable Fermi coupling constant $G_{F}$ is replaced by the $%
W^{\pm }$ propagators (see, for a detailed pedagogical exposition, \cite%
{We96})
\begin{equation*}
G_{F}\rightarrow \frac{g_{W}^{2}}{M_{W}^{2}-q^{2}},
\end{equation*}
where $g_{W}$ is the (dimensionless) electro-weak coupling constant and $%
M_{W}$ is the mass of the $W$-bosons. In the gravitational case, the large
\textbf{N} expansion suggests that, in the strongly coupled phase of
gravity, a similar phenomenon should occur:
\begin{equation*}
G\rightarrow G\frac{M^{2}}{M^{2}-\Pi (q^{2})}.
\end{equation*}
Thus, the sentence \textquotedblright gravity is an effective field
theory\textquotedblright\ would acquire a rather precise meaning. Such a non
renormalizability should be removed by non perturbative effects, displayed
by the large \textbf{N} expansion, which correctly take into account the
heavy degrees of freedom in a way very similar to what happens in the
electro-weak model.

\section{Possible complications}

The main effects which have been neglected in the previous discussion are
related to the ghosts. In particular, ghosts loops could cancel out,
asymptotically, the $B$ loops vanifying the previous UV softening. In fact,
the large \textbf{N} expansion itself provides one with a natural recipe to
deal with this problem.

\begin{figure}[tbp]
\begin{center}
\includegraphics*[scale=.30,angle=0]{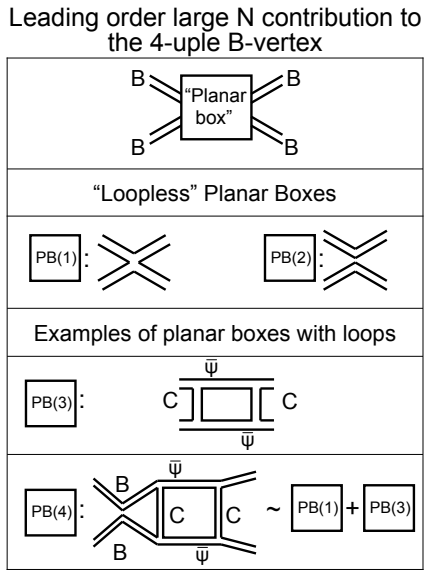}
\end{center}
\caption{In this picture the double line structures of the possible
contributions to the 4-uple $B$ vertex have been drawn. Only two structures
could contribute at the \textit{tree level large} \textbf{N} 4-uple $B$
vertex: PB(1) (which, however, is not present in the theory) and PB(2). All
others planar boxes with four external $B$ lines have at least one closed
color loop inside: therefore, they are not relevant as far as UV-power
counting is concerned.}
\label{7b}
\end{figure}

In the gravitational case, due to the \textquotedblright higher
spin\textquotedblright\ Lagrange multiplier $\phi _{ab}$, it is possible to
construct planar diagrams without closed color loops (thus contributing to
the large \textbf{N} bare propagators) containing an infinite number of
loops of the standard Feynman expansion\footnote{%
As it has been already remarked, this feature tells far apart gravity from
gauge theories: in gravity the bare large \textbf{N} propagators and
vertices do not coincide with the standard bare propagators and vertices in
the Feynman expansion.}. The question is: do the ghosts contributions to the
4-uple $B$-vertex survive at the tree-level in the large \textbf{N }%
expansion in such a way to prevent the above analyzed UV softening? In other
words, is it possible to construct a contribution to the 4-uple $B$-vertex
by using the ghosts vertices in Eqs. (\ref{vert34}), (\ref{vert56}), (\ref%
{vert78}), (\ref{vert910}) and (\ref{vertu11}) \textit{without closed
internal lines}? The key point is that the only way in which ghosts effects
could vanify the previously considered UV softening is by changing the
tree-like large \textbf{N} vertices and, in particular, the 4-uple $B$
vertex (since the large \textbf{N} power counting only depends on tree-like
large \textbf{N} vertices and propagators): ghosts effects have to come into
play at tree-level otherwise the power-counting does not change. It is easy
to see that all the possible ghosts-mediated corrections to the 4-uple $B$
vertex contain at least one closed color loop (since the vertices in Eqs. (%
\ref{vert34}), (\ref{vert56}), (\ref{vert78}), (\ref{vert910}) and (\ref%
{vertu11}) have connected structures) and, therefore, do not contribute to
the bare large \textbf{N} propagators and vertices. Let us consider the
ghost contribution to the $B$ vertex in Fig.~\ref{6b}: indeed, at genus
zero, it contains at least one closed color loops. More in general, in the
presence of fields living in the adjoint representation of the gauge group
and vertices with the ghosts vertices of the BF formulation of gravity it is
not possible to construct \textquotedblright tree-like large \textbf{N}%
\textquotedblright\ diagrams without closed color loops contributing to the
4-uple $B$ vertex (see Fig.~\ref{7b}). Therefore, ghosts effects do not
affect the UV softening of the previously considered \textquotedblright
bubble-like\textquotedblright\ series.

\subsection{The role of $\protect\eta $}

Up to now, the role of $\eta $\ has not been discussed yet. Its introduction
(as in the Yang-Mills case) is dictated by technical reasons. To deduce the
large \textbf{N} effects on the interactions of $\eta $\ and $B$ one can
consider the large \textbf{N} improvements for the 4-uple vertex without
separating $B^{\prime }$ in $B$ and $\nabla \eta $. Obviously, since $B$ and
$\nabla \eta $ can be \textquotedblright assembled\textquotedblright\ back
in a unique field, the vertices between $\nabla \eta $ and $B$ experience at
large \textbf{N} similar UV improvements as in Eq. (\ref{ultimpropro1}). In
other words, $\eta $\ represents the longitudinal components of $B^{\prime }$
so that the separation of $B^{\prime }$ in $B$ and $\nabla \eta $ is an
artifact, the physical field is $B+\nabla \eta $. This implies that one can
trivially deduce the large \textbf{N} effects on the interactions between $%
\nabla \eta $\ and $B$ by analyzing the large \textbf{N} improvements of the
effective 4-uple vertex for the physical field $B+\nabla \eta $\footnote{%
Which is given by the already discussed bubble-like geometric series
\textquotedblright mediated\textquotedblright\ by the \textquotedblright
higher spin\textquotedblright\ Lagrange multiplier $\phi _{ab}$ which allows
\textquotedblright bubble-like\textquotedblright\ series with the same
internal-line structure in Fig.\ref{4}. Thus, the 4-uple vertex of physical
field $B+\nabla \eta $ is dressed by the same factor which has been
discussed in the previous section.} and then separating it into $B$ and $%
\nabla \eta $.\ On the other hand, since $\eta $ can be gauged to zero
(being one of its transformation laws given by a shift) its properties do
not affect the UV behavior.

As it has been already mentioned, at this stage of the analysis, it is not
clear if and how one could avoid the introduction of the auxiliary $\eta $\
field from the very beginning. Indeed, the introduction of $\eta $ and of
the \textquotedblright Yang-Mills\textquotedblright\ term\footnote{%
The introduction of the Yang-Mills term has not played any explicit role in
the whole discussion. Its only role is, by the way, to arrive at a theory
which has the same propagators of the BF Yang-Mills theory.} is not a mere
avoidable technical complication: it is in fact the more transparent way to
identify the \textquotedblright guilty\textquotedblright\ vertex and to cure
it by the means of the large \textbf{N} expansion.

\section{A possible physical interpretation}

It will be now discussed a possible interpretation of the previous results
in terms of a sort of gravitational confinement (in which the gravitational
color, that is the spin, is confined) which should be dual, in the sense of
the Gauge/Gravity correspondence (see, for two detailed reviews, \cite%
{Ber03,Di05}) to the standard gauge theoretical confinement.

In many other cases, such as the $O(\mathbf{N})$ scalar (in five space-time
dimensions) and Fermionic (in three space-time dimensions) models with
quartic interactions, the seemingly \textquotedblright
magic\textquotedblright\ properties of the large \textbf{N} resummations
are, in fact, related in a very simple way to physical properties of the
models. In the above mentioned cases, the large \textbf{N} expansion is able
to explore the strongly coupled phase of the theory in which phenomena such
as the appearance of \textquotedblright color-less\textquotedblright\ bound
states in the spectrum, spontaneous breaking of symmetries and so on occur.
Thus, a natural question is: which is the phenomenon behind the improvement
of the UV behavior of gravity at large \textbf{N}? In the present
formulation of the gravitational action matter fields should be described as
scalar fields living in suitable representation of the internal gauge group
according to their spin \cite{Ca05}. At large \textbf{N}, in the strongly
coupled phase of gravity (as it is also suggested by the Gauge/Gravity
correspondence), the physical spectrum should be dominated by colorless
bound states. In this context colorless particles simply means scalar
particles which, therefore, should dominate the spectrum in the UV region%
\footnote{%
Gravitational confinement would be of great cosmological importance since it
would provide the standard inflationary scenario with sound basis: such a
mechanism could lead to the expected decrease of the number of degrees of
freedom expected in a holographic theory.}. Gravitational confinement would
dramatically improve the renormalizability of the theory. The reason is
that, if the gravitational interaction confines at the Planck scale, one is
left with only scalar fields at high energy: this would soften the UV
behavior of the theory. This can be seen forgetting for a moment the present
formalism (in which the spin of particles is represented by an internal
index in a suitable representation of the internal gauge group) and
returning to the standard \textquotedblright space-time\textquotedblright\
interpretation of spinful particles. The contribution to the scattering
amplitude $A_{J}(s,t)$ in the $t-$channel (where $t$ and $s$ are Mandelstam
variables) at high energies of a particles with spin equal to $J$ is roughly
\begin{equation*}
A_{J}(s,t)\approx -e^{2}\frac{s^{J}}{t-m_{J}^{2}},
\end{equation*}
$m_{J}$ being the mass of the particle and $e$ being a suitable coupling
constant. When $J<1$, loops integrals in which such a spin $J$ particle
appears are convergent; when $J=1$ there are logarithmic divergences (which,
in case, can be renormalized) and when $J>1$ there are non renormalizable
divergences. This fact is behind the perturbative non renormalizability of
Einstein-Hilbert action. In fact, if there is a critical energy scale,
beyond which only scalars are left in the spectrum due to the gravitational
confinement, the above problems are naturally solved by the dynamics of the
gravitational field: thus, the gravitational interaction could be non
perturbatively renormalizable.

It is worth to note that the main responsible for the good UV properties of
gravity at large \textbf{N} is, presumably, a sort of gravitational
confinement. On the other hand, in gauge theories, suitable order parameters
for confinement are the Wilson loops. Therefore, one should expect that the
gravitational analogue of the Wilson loops should play a fundamental role in
understanding the strongly coupled phase of gravity. Interestingly enough,
such operators have been introduced in \cite{RS88} in the context of the
Ashtekar formalism for gravity \cite{As86} (for an updated review, see \cite%
{As04}). Thus, it is apparent an intriguing relation between loop quantum
gravity and the present large \textbf{N} expansion for gravity.

This is a good point to think at the following question: what would be the
physical meaning of renormalizability at large \textbf{N}? As far as the
standard perturbative expansion is concerned, a pragmatic answer is that a
given theory (without local symmetries) is renormalizable (in the standard
Feynman expansion) when the number of primitive divergent Feynman diagrams
is finite. This issue can be dealt with by looking at the UV behavior of
bare propagators and vertices (on which the standard perturbative expansion
is based). The deep physical meaning of such a property is manifest in the
Wilson point of view (to discuss the Wilson view in a gravitational context
is far beyond the scope of this paper). As far as the present scheme is
concerned, it is enough to say that an UV renormalizable theory is a theory
which allows meaningful and predictive computations (in the UV) because the
fields appearing in the Lagrangian are suitable to describe the UV phase%
\footnote{%
That is, the fields in the Lagrangian represents small fluctuations around
the UV vacuum so that the perturbation expansion works.}. When this does not
happen, there are two possibilities: either the theory is wrong or the UV
degrees of freedom are non trivial non local combinations of the ones
appearing in the Lagrangian\footnote{%
For instance, in QCD quarks and gluons are good UV degrees of freedom but
they are not good IR degrees of freedom. This, of course, does not imply
that QCD is wrong: it is an indication that the IR degrees of freedom are
non trivial combinations of the UV degrees of freedom.}.

Thus, a pragmatic answer to the question on the physical meaning of
renormalizability at large \textbf{N}\footnote{%
Large \textbf{N} renormalizability (in the cases in which there are not
local symmetries) can be "detected" as standard renormalizability using
power counting arguments by looking at the UV behavior of large \textbf{N}
propagators and vertices (that is, vertices and propagators which have been
corrected in order to encode the leading corrections in the expansion).}
could be the following: a theory which is UV renormalizable at large \textbf{%
N} (and is not UV renormalizable in the standard perturbative expansion such
as the Gross-Neveu model in 3 dimensions or the $\phi ^{4}$ model in 5
dimensions \cite{Pa75}) is a theory which is formulated in terms of fields
not suitable to describe the UV phase. However, it is not wrong: the large
\textbf{N} recipe tells how to sum class of diagrams of the standard
expansion in order to obtain a meaningful expansion (with a finite number of
primitive superficially divergent diagrams). In the already mentioned
examples, this has a very precise meaning: in the UV phase new degrees of
freedom appear which are bound states of the original fields (see, for
instance, \cite{MZ03}). This is exactly the picture which seems to emerge
from the present scheme: scalar bound states which soften the UV behavior of
amplitudes of particles carrying spin 2 or greater. However, as it will be
discussed in a moment, in gravity there is the further complication of local
symmetries: one should also prove that local symmetries are preserved by the
large \textbf{N} resummations.

\subsection{Connections with the KLT relations}

\begin{figure}[tbp]
\begin{center}
\includegraphics*[scale=.30,angle=0]{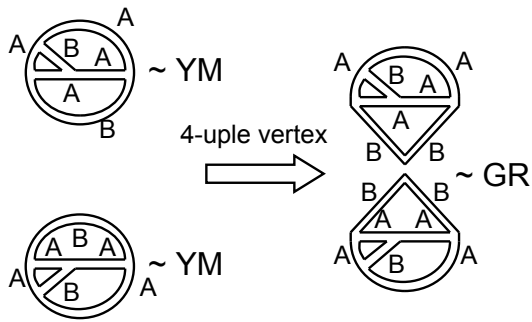}
\end{center}
\caption{{}In this picture a typical gravitational large \textbf{N} diagram
which could be responsible for the KLT relations has been drawn. It is
manifest the important role which could have the 4-uple $B$ vertex (which is
disconnected from the "internal lines" point of view) in explaining such
important relations between gravity and gauge theory (which, in the BF
formalism, are distinguished precisely by the above mentioned 4-uple $B$
vertex).}
\label{9}
\end{figure}

It is worth to mention here an interesting relation (which is worth to be
further investigated) with the so called \textit{KLT} relations \cite{KLT86}%
. The \textit{KLT} relations were first obtained in a string theoretical
framework: they allow to express closed string amplitudes in terms open
string amplitudes. Schematically, the factorization of the vertex operators
\begin{equation*}
V^{closed}=V_{left}^{open}\times \overline{V}_{right}^{open}
\end{equation*}%
(where the $V$s are vertex operators of the closed string, open string left
modes and right modes) is related to the results in \cite{KN69} stating that
correlations of vertex operators factorize at the level of integrands (that
is, before the world sheet integrations are performed). Kawai, Lewellen and
Tye were able to proof a stronger result: the complete closed string
amplitudes factorize into products of open string amplitudes even after the
world sheet integrations. Indeed, the \textit{KLT} relations go well beyond
string theory itself: they imply, for instance, in the low energy limit
highly non trivial relations among tree amplitudes of gravity in four
dimensions which factorize (in suitable non linear gauges) into gauge
theoretical tree amplitudes in four dimensions. These results have been
generalized to include loops by using unitarity relations (for a pedagogical
review see \cite{Be02}), however it is not available yet a complete proof of
these very useful factorizations results. Usually, one deals with the KLT
relations in the standard metric formalism in which, having in mind small
deviations from a flat metric
\begin{equation*}
g_{\mu \nu }\sim \eta _{\mu \nu }+h_{\mu \nu },
\end{equation*}%
the basic variable is the metric fluctuation $h_{\mu \nu }$. In the standard
metric formulation of gravity the structure of the many vertices present in
the theory is very complicatet. To fully exploit the KLT relations one needs
to choose a gauge in which, roughly speaking, the right index and the left
index of $h_{\mu \nu }$ are never contracted with eachothers (otherwise no
"left-right" factorization
\begin{equation*}
h_{\mu \nu }\sim \epsilon _{\mu }^{-}\otimes \epsilon _{\nu }^{+}
\end{equation*}%
would be apparent\footnote{$\epsilon _{\mu }^{\pm }$ have to be thought as
the "gluons" of the gauge theory responsible of the factorization of the
gravitational amplitudes (the plus and minus signs refer to the elicities).}%
). Such (non linear) gauge choices are highly non trivial and the many
vertices of the theory in the metric formalism sometimes obscure the
physical origin behind the KLT relations. In the present framework such
features are rather manifest: in particular, the presence of a Lagrange
multiplier field $\phi _{ab}$\ which, in the double line notation, is
represented by four internal lines leads directly to amplitudes fulfilling
the generalized KLT relations. The reason is that in the BF formulation of
gravity the propagators and vertices can be chosen to be equal to the
propagators and vertices appearing in the BF formulation of Yang-Mills
theory: the only, crucial, difference is the higher spin Lagrange multiplier
field. Such a field gives rise, quite generically, to gravitational
amplitudes which manifestly are factorized into pieces in which only "YM
fields" (that is, fields which are also present in the BF Yang-Mills
Lagrangian) appear (see, for instance, Fig.\ref{9}). In other words, $\phi
_{ab}$\ "allows" to attach amplitudes of the BF Yang-Mills theory to obtain
amplitudes of the BF formulation of gravity: for this reason, the present
scheme seems to be suitable to fully exploit and, hopefully, to establish in
general the KLT relations.

\section{Conclusions and Perspectives}

In this paper an analysis of renormalizability of gravity at large \textbf{N}
expansion for General Relativity has been carried on. It is based on the BF
formulation of General Relativity in which the Einstein-Hilbert action is
splitted into a topological term plus a constraint. It has been shown that
the large \textbf{N} expansion dictates resummations of a suitable class of
planar diagrams which lead to a great improvement of the UV behavior of
gravity: only a finite number of superficially divergent diagrams are
present at large \textbf{N}. This is an important step in proving
renormalizability. The next steps are the analysis of the infinities and
their fulfilment of the symmetry constraints such as the Slavnov-Taylor
identities and the Zinn-Justin equation. The analysis of the KLT relations
in this scheme is also worth to be further investigated.

%\begin{ack}
\section*{Acknowledgements}
The author would like to thank Prof. G. Vilasi for continuous
encouraging, P. Vitale for important bibliographic suggestions and
L. Parisi for invaluable help in drawing the pictures. This work
has been partially supported by PRIN SINTESI 2004.
%\end{ack}

\end{document}